\documentclass{pos}

\usepackage{pstricks}
\usepackage{color}
\let\normalcolor\relax

\title{Determination of the strong coupling constant from inclusive jet cross section in $p\bar{p}$ collisions at $\sqrt{s}=1.96$~TeV with the D\O\ experiment }

\ShortTitle{Determination of $\alpha_s$ from D\O\ inclusive jet cross section}

\author{Lars Sonnenschein \\ %\thanks{A footnote may follow.}\\
        RWTH Aachen University \\
        III. Phys. Inst. A \\
        52056 Aachen \\
        Germany \\
        E-mail: \email{Lars.Sonnenschein@cern.ch} \\
       {\bf on behalf of the D\O\ collaboration}
    }

\abstract{
The strong coupling constant $\alpha_s$ and its dependence on the momentum scale is determined 
from the $p_T$ dependence of the inclusive jet cross section in $p\bar{p}$ collisions at 
$\sqrt{s}=1.96$~TeV measured with the D\O\ experiment. The jet transverse momentum range of 
$50 < p_T < 145$~GeV contributes to the determination. Using perturbative QCD caclulations to 
order ${\cal{O}}(\alpha_s^3)$ combined with resummed threshold corrections to order 
${\cal{O}}(\alpha_s^4)$ an $\alpha_s(M_Z)=0.1161^{+0.0041}_{-0.0048}$ is obtained. This is the 
most precise result from a hadron-hadron collider.
}

\FullConference{XVIII International Workshop on Deep-Inelastic Scattering and Related Subjects, DIS 2010\\
                April 19-23, 2010\\
                Firenze, Italy}

\begin{document}

\section{Introduction}

A remarkable property of quantum chromodynamics (QCD) is the property of asymptotic freedom. 
It is reflected by the prediction of the renormalisation group equation (RGE) dependence of 
the strong coupling constant $\alpha_s$ on the momentum scale. Experimental tests of asymptotic 
freedom require precise determinations of $\alpha_s$ over a large range of momentum scales. Its 
determination discussed here~\cite{prdas}\cite{d0as} is based on an inclusive jet cross section 
measurement~\cite{d0inc} with the D\O\ detector~\cite{d0det}.
The inclusive jet cross section $d^2\sigma_{\mbox{\scriptsize jet}}/dp_Td|y|$ has been measured
using the Run~II iterative midpoint cone ($R=0.7$) algorithm~\cite{d0cone} in the energy scheme.
110 data points are measured as a function of the momentum transverse to the beam axis, $p_T$
in six %disjoint adjacent 
\begin{figure}[t]
\vspace*{-3ex}
%\centerline{
\unitlength 1cm
\begin{picture}(9.0,6.3)
\put(3.5,-0.2){\includegraphics[width=0.66\columnwidth]{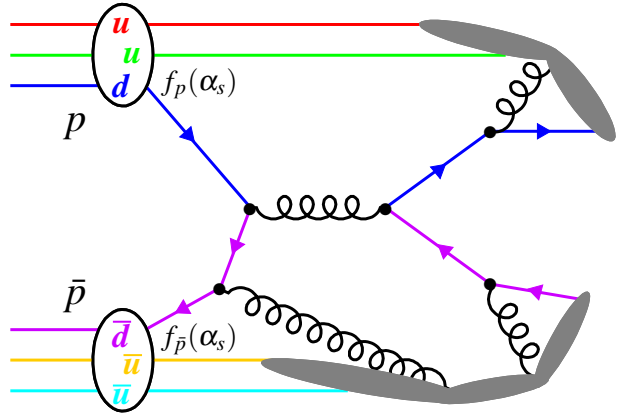}}
\put(4.2, 4.2){\scalebox{1.4}{$p$}}
\put(4.2, 1.9){\scalebox{1.4}{$\bar{p}$}}
\put(5.5, 4.75){\scalebox{1.0}{$f_p(\alpha_s)$}}
\put(5.5, 1.35){\scalebox{1.0}{$f_{\bar{p}}(\alpha_s)$}}
\end{picture}
\vspace*{-4ex}
\caption{A drawing, illustrating jet production in a proton anti-proton collision 
with the hard scattering process, initial state, final state radiation and hadronisation 
(jet fragmentation) including the underlying event. 
}\label{Fig:d0escalefeyn}
\vspace*{-1ex}
\end{figure}
equidistant regions of $|y|$ (absolute rapidity) spanning together a range of $0<|y|<2.4$.

The initial state of an inelastic $p\bar{p}$ collision is given by non-perturbative Parton 
Distribution Functions (PDF's) $f$ which are determined empirically and follow a momentum 
scale evolution according to QCD. The partonic final state
is determined by a convolution of perturbative QCD (pQCD) scattering amplitudes and the PDF's. 
The cross section for the hard scattering process of partons can be calculated via a perturbative
expansion in orders of $\alpha_s$ as
\begin{equation}
  \sigma_{\mbox{\scriptsize pert.}(\alpha_s)} 
  = \sum_n \left( \alpha_s^nc_n \right) \otimes f_p(\alpha_s) \otimes f_{\bar{p}}(\alpha_s) \; .
\end{equation}
The pQCD calculations are corrected for non-perturbative effects of hadronisation and underlying 
event (see Fig. \ref{Fig:d0escalefeyn}). The prediction of the theory can then be expressed as
\begin{equation}
  \sigma_{\mbox{\scriptsize theory}}(\alpha_s) = \sigma_{\mbox{\scriptsize pert.}}(\alpha_s) \cdot c_{\mbox{\scriptsize non-pert.}} \; ,  
\end{equation}
where the factor $c_{\mbox{\scriptsize non-pert.}}$ takes into account the non-perturbative 
corrections.
At the same time measurements are corrected for detector effects. Comparisons
between measurements and predictions of theory are then accomplished at the 
hadronic final state \cite{candl}.

Previous determinations of $\alpha_s$ by means of hadronic jet cross section measurements have 
had large uncertainties in comparison to other determinations. They have not been very 
competitive but now the tools for substantial improvements are available. More precise 
theory calculations together with more and better measured data lead to smaller uncertainties 
and competitive results. The jet energy scale has been continuously improved over many years
and reaches at present a fractional uncertainty of $1 - 2\%$ over a wide range of jet 
transverse momenta from 50~GeV to 600~GeV.

Tevatron jet cross section measurements cover a wide kinematic range.
In the kinematic plane of momentum fraction $x$ and scale $Q^2$ unique regions of phase 
space are covered, in particular when forward jets up to absolute values of pseudorapidity 
of three are included. There is also overlap with measurements from $ep$ collisions.
Therefore the D\O\ data allow precision tests of pQCD and an independent extraction of $\alpha_s$.

\section{Extraction of $\alpha_s$}

The used perturbative calculations are a sum of a full calculation to order 
${\cal{O}}(\alpha_s^3)$ $\left[\right.$ next-to-leading order (NLO)$\left.\right]$ combined with 
${\cal{O}}(\alpha_s^4)$ resummation of (2-loop) terms from threshold corrections \cite{kidonakis}. 
Adding the 2-loop threshold corrections yields a significant reduction of renormalisation and 
factorisation scale dependence which in turn leads to significantly smaller uncertainties in the 
theoretical calculations. These calculations have been ignored for a long time and became only
popular and accessible since they are included into fastNLO \cite{fastNLO} which
provides fast recalculations for arbitrary PDF's and is based on 
NLOJET++ \cite{znagy03} \cite{znagy02} and code from authors of ref. \cite{kidonakis}.
The calculations are performed in the $\overline{\mbox{MS}}$ scheme
\cite{bardeen} for five active quark flavours  using the next-to-next-to-leading logarithmic
(3-loop) solution to the RGE. 
The PDF's are taken from the MSTW2008 next-to-next-to-leading order (NNLO) parameterisations
\cite{mstw09a} \cite{mstw09b}. The jet transverse momentum is taken for the renormalisation 
and factorisation scale.

The determination of $\alpha_s$ is accomplished from inclusive jet cross section data points
by minimising a $\chi^2$ function between data and theory. The $\alpha_s(p_T)$ values are evolved
via the 3-loop solution to the RGE to $\alpha_s(M_Z)$. The central $\alpha_s(M_Z)$ result is
obtained by minimising $\chi^2$ with respect to  $\alpha_s(M_Z)$ and nuisance parameters for
correlated uncertainties.

The procedure described above to determine $\alpha_s$ requires the knowledge of 
$\sigma_{\mbox{\scriptsize pert.}}(\alpha_s(M_Z))$ as a continuous function of $\alpha_s(M_Z)$
over a $\alpha_s(M_Z)$ range which covers possible fit results and uncertainties.
The MSTW2008 NNLO (NLO) parameterisations are available for 21 different values  of
$\alpha_s(M_Z)$ in a range of 0.107 - 0.127 (0.110 - 0.130) in steps of 0.001.
CTEQ 6.6 parameterisations \cite{cteq6.6} are available up to NLO for 5 different values of $\alpha_s(M_Z)$.
The MSTW2008 PDF's with fine spacing in $\alpha_s(M_Z)$ over a wide range are used
together with cubic spline interpolations for the main result. 

The D\O\ inclusive jet production cross section measurement \cite{d0inc} 
entered already in the \linebreak
MSTW2008 PDF's.
The correlations between experimental and PDF uncertainties are not documented.
\begin{figure}[bh]
%\centerline{
\unitlength 1cm
\begin{picture}(9.0,6.85)
\put(0.2,0.1){\includegraphics[width=0.463\columnwidth]{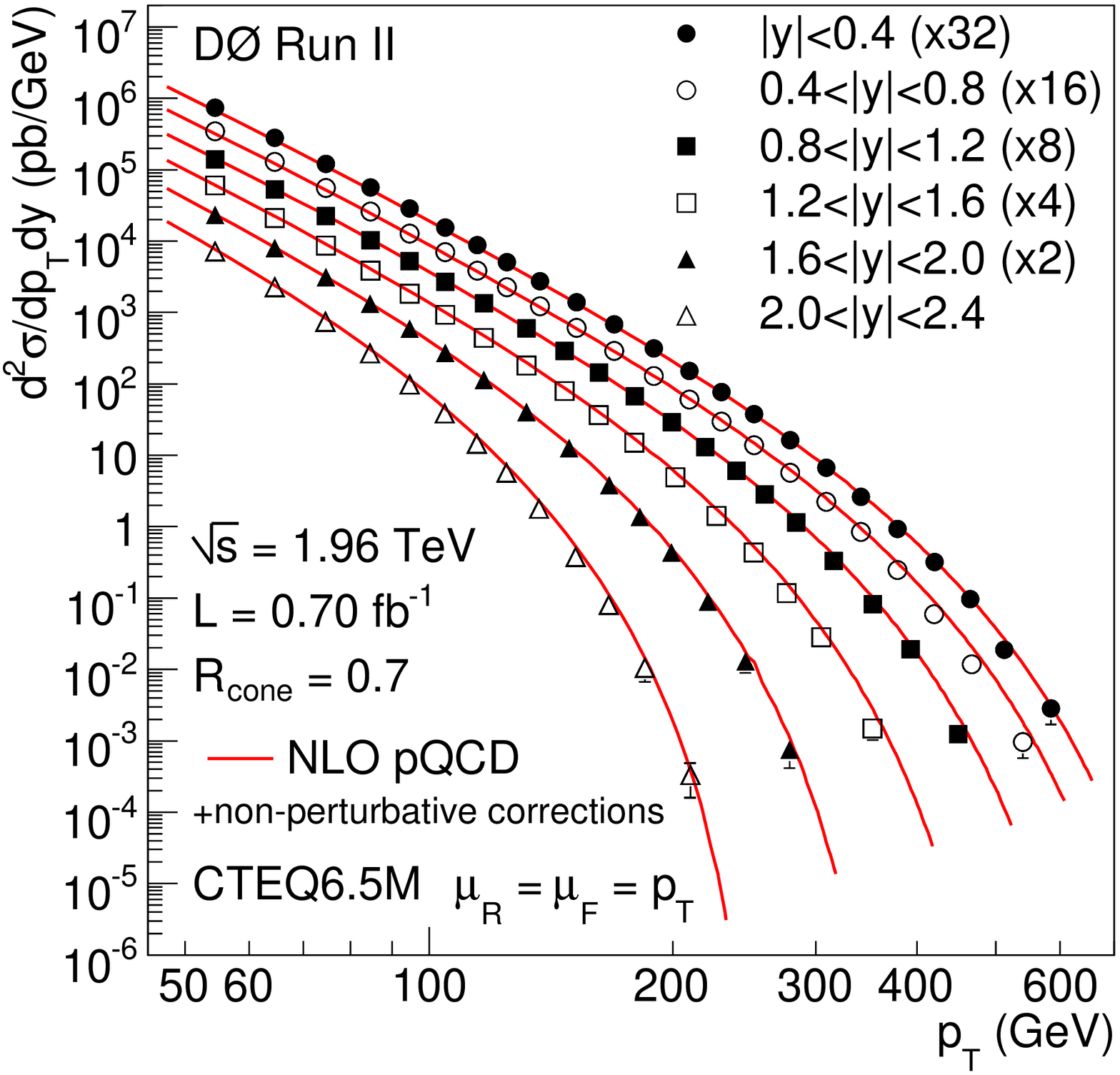}}
\put(1.34,6.29){\color{blue}\rotatebox{-27.5}{\line(1,0){2.57}}}
\put(3.62,5.115){\color{blue}\rotatebox{-90}{\line(1,0){0.28}}}
\put(1.34,6.29){\color{blue}\rotatebox{-90}{\line(1,0){0.74}}}
\put(1.34,5.55){\color{blue}\rotatebox{-27.9}{\line(1,0){0.797}}}
\put(2.04,5.183){\color{blue}\rotatebox{90}{\line(1,0){0.202}}}
\put(2.04,5.38){\color{blue}\rotatebox{-28.7}{\line(1,0){0.62}}}
\put(2.59,5.07){\color{blue}\rotatebox{90}{\line(1,0){0.202}}}
\put(2.59,5.264){\color{blue}\rotatebox{-28.9}{\line(1,0){0.72}}}
\put(3.211,4.92){\color{blue}\rotatebox{90}{\line(1,0){0.16}}}
\put(3.211,5.08){\color{blue}\rotatebox{-30.3}{\line(1,0){0.48}}}

\put(8.2, 0.0){\includegraphics[width=0.44\columnwidth]{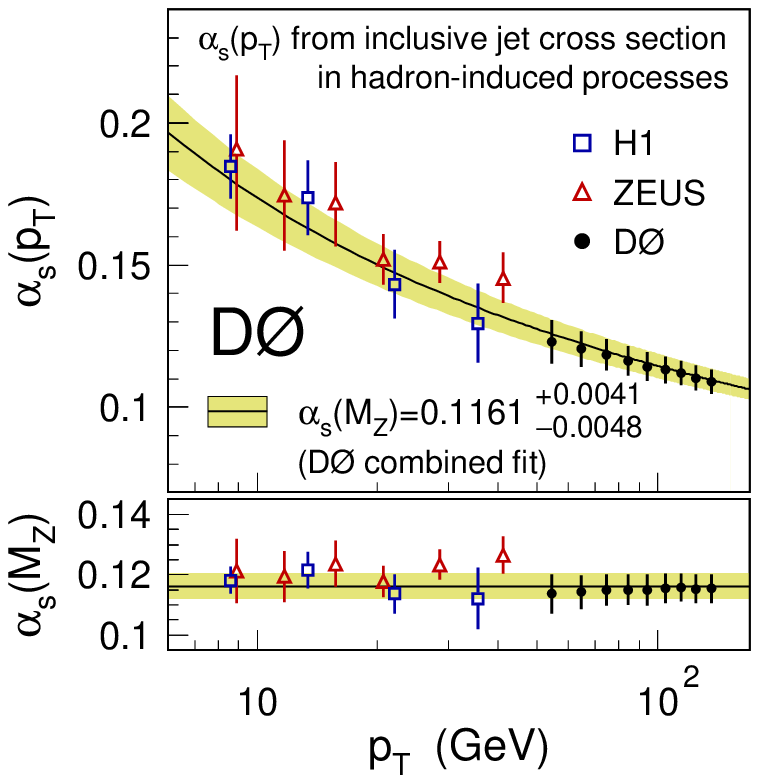}}

\end{picture}
\vspace*{-2.5ex}
\caption{Left: Inclusive double differential jet production cross
section measured with the D\O\ detector \cite{d0inc}. 
The 22 selected data points for the extraction of $\alpha_s$ are indicated 
by the surrounding blue polygon. \newline
Right: The strong coupling constant $\alpha_s$ as a function of 
$p_T$ (top) and at $M_Z$ (bottom). 
For comparison HERA DIS jet data are superposed.
Error bars correspond to total uncertainties.
}\label{Fig:d0incsel}
%\end{wrapfigure}
\vspace*{-3ex}
\end{figure}
Therefore they can not be taken into account in the extraction of $\alpha_s$ and
the cross section data points which provide strong constraints on the PDF's have to be avoided.
Precision Tevatron jet data dominate models for gluon densities at high momentum fraction $x$ 
and start to affect gluon densities at $x\gtrsim 0.25$.
At leading order, di-jet events access $x$ values of $x_a=x_T\frac{e^{y_1}+e^{y_2}}{2}$ and
$x_b=x_T\frac{e^{-y_1}+e^{-y_2}}{2}$ where $x_T$ is defined by $x_T=\frac{2p_T}{\sqrt{s}}$.
The mapping is less unique in inclusive jet production cross section data where 
the $x$-value is not fully 
constrained given a measured bin of $p_T$ and $|y|$. The full kinematics is unknown since 
it is integrated over the number of jets. The momentum fraction can be approximately
constrained in treating the sub-leading jets as central ($|y|=0$). The requirement 
$\tilde{x}<0.15$ removes all data points for which more than half the cross section is produced
at $x_{\max}\gtrsim 0.25$. 22 out of the 110 data points fulfil this requirement (see 
Fig. \ref{Fig:d0incsel}, left).
They are distributed in the kinematic phase space as follows:
nine data points in the interval $|y|<0.4$, $50<p_T<145$~GeV;
seven data points in the interval $0.4<|y|<0.8$, $50<p_T<120$~GeV;
four data points in the interval $0.8<|y|<1.2$, $50<p_T<90$~GeV and
two data points in the interval , $50<p_T<70$~GeV.

The pQCD uncertainties due to uncalculated higher orders are estimated in varying the
renormalisation and factorisation scales a factor of two up and down.
In the kinematic region used for the extraction of $\alpha_s$ the two scales show positively
correlated effects on the cross sections. As a conservative estimate both scales are varied
into the same direction simultaneously. The resulting uncertainties can not be treated as 
Gaussian. Therefore the $\chi^2$ fit is repeated for the scale variations and the differences
with respect to the central value of $\alpha_s(M_Z)$ are added in quadrature to the other 
uncertainties.

\section{Results}

A combined fit to the 22 selected data points, grouped in nine transverse momentum
intervals, yields $\alpha_s(M_Z)=0.1161^{+0.0041}_{-0.0048}$.
The results are shown in Fig. \ref{Fig:d0incsel} (right) 
as nine $\alpha_s$ values in the range of 
$50 < p_T < 145$~GeV with their total uncertainties which are largely correlated among 
the data points. The results are complemented by DIS jet data from HERA. The $\alpha_s(p_T)$ 
results obtained here are consistent with the energy dependence predicted by the RGE.
The combined result for $\alpha_s(M_Z)$ is consistent with the combination of HERA jet 
data~\cite{hera} and the world average~\cite{bethke}.
The largest contribution to the total uncertainty comes from the correlated experimental 
uncertainties which are dominated by jet energy calibration, jet transverse momentum 
resolution and integrated luminosity.

Variation of the non-perturbative uncertainties by a factor of two up and down
changes the central value by +0.0003 and -0.0010. It does not affect the uncertainty of 
the combined $\alpha_s(M_Z)$ result. Replacing the MSTW2008 NNLO PDF's by the CTEQ 6.6 PDF's
increases the central result by only +0.5\%, which is much less than the PDF uncertainty.
Excluding the 2-loop threshold corrections and using pure NLO pQCD together with the MSTW2008
NLO PDF's and the 2-loop solution to the RGE yields $\alpha_s(M_Z)=0.1201^{+0.0072}_{-0.0059}$.
The small increase in the central value originates from the missing ${\cal{O}}(\alpha_s^4)$
contributions which are compensated by a corresponding increase of $\alpha_s$. 
The difference to the central result is well within the scale uncertainty of the NLO result.

\section{Conclusions}
The strong coupling constant has been determined from the D\O\ inclusive jet cross section 
measurement~\cite{d0inc} using theory prediction at NLO with the resummation of 2-loop terms 
from threshold corrections, i.e. at NNLO accuracy. The $\alpha_s(p_T)$ results are consistent 
with the predicted energy dependence of the renormalisation group equation. The combined 
result from 22 selected data points is $\alpha_s(M_Z)=0.1161^{+0.0041}_{-0.0048}$. This is the 
most precise result at a hadron collider.

%%%%%%%%%%%%%%%%%%%%%%%%%%%%%%%%%%%%%%%%%%%%%%%%%%%%%%%%%%%%%%%%%%%%%%%%%%%%%%%%%%%%

%\section{Acknowledgments}

%Many thanks to the staff members at Fermilab and collaborating institutions. 
%This work has been supported by the DOE and NSF (USA); CEA and CNRS/IN2P3 (France); 
%FASI, Rosatom and RFBR (Russia);
%CNPq, FAPERJ, FAPESP and FUNDUNESP (Brazil); DAE and DST (India); Colciencias (Columbia); 
%CONACyT (Mexico); KRF and KOSEF (Korea); CONICET and UBACyT (Argentina); FOM (The Netherlands);
%STFC (United Kingdom); MSMT and GACR (Czech Republic); CRC Program, CDF, NSERC and WestGrid Project 
%(Canada); BMBF, DFG and the Alexander von Humboldt Foundation (Germany); SFI (Ireland); 
%The Swedish Research Council (Sweden); and CAS and CNSF (China).

%\section{Bibliography}

% ****************************************************************************
% BIBLIOGRAPHY AREA
% ****************************************************************************

\begin{footnotesize}
% IF YOU DO NOT USE BIBTEX, USE THE FOLLOWING SAMPLE SCHEME FOR THE REFERENCES
% ----------------------------------------------------------------------------

% ----------------------------------------------------------------------------

\end{footnotesize}

% ****************************************************************************
% END OF BIBLIOGRAPHY AREA
% ****************************************************************************

\end{document}